\documentstyle[12pt,epsfig]{article}

\newcommand{\be}{\begin{eqnarray}}
\newcommand{\ee}{\end{eqnarray}}

\def\nue{{\nu_e}}
\def\anue{{\bar\nu_e}}

\newcommand{\ms}{\Delta m^2_{21}}
\newcommand{\ma}{\Delta m^2_{31}}

\newcommand{\sss}{\sin^2 \theta_{12}}
\newcommand{\sch}{\sin^2 \theta_{13}}

\newcommand{\chisq}{$\chi^2$}
\newcommand{\taet}{T_\anue{\rm (true)}}
\newcommand{\txt}{T_x{\rm (true)}}
\newcommand{\sig}{$3\sigma$}
\def\ltap{\ \raisebox{-.4ex}{\rlap{$\sim$}} \raisebox{.4ex}{$<$}\ }

\begin{document}

\begin{flushright}
OUTP-0509P\\
SINP/TNP/05-26
\end{flushright}
\bigskip

\begin{center}
{\Large \bf 
Possible violation of the spin-statistics relation for neutrinos: 
checking through future galactic supernova}

\vspace{.5in}

{\bf 
Sandhya Choubey$^{\star a}$ and Kamales Kar$^{\dagger b}$}
\vskip .5cm
$^a${\it Rudolf Peierls Centre for Theoretical Physics,}
{\it University of Oxford,}\\
{\it 1 Keble Road, Oxford OX1 3NP, UK}\\
\vspace{.5cm}
$^b${\it Theory Group, Saha Institute of Nuclear Physics,}\\
{\it 1/AF, Bidhannagar,
Calcutta 700 064, India}

\vskip 1in

\end{center}

\begin{abstract}
We use the detection of neutrinos from a future galactic type-II
supernova event in a water Cerenkov detector like Super-Kamiokande
to constrain the possible violation of spin-statistics by 
neutrinos resulting in their obeying a mixed statistics instead of 
Fermi-Dirac.
\end{abstract}

\vskip 2in

\noindent $^\star$ email: sandhya@thphys.ox.ac.uk

\noindent $^\dagger$ email: kamales.kar@saha.ac.in

\newpage

Recently there was a suggestion of neutrinos violating the spin-statistics
relation and thereby becoming a good candidate for all (or part) of the dark
matter in the universe \cite{DS}. 
There are a number of papers which discuss the
possible violation of spin-statistics relation by neutrino though no 
consistent
satisfactory model exists \cite{spin1,spin2,spin3,spin4}. 
On the other hand the experimental verification of neutrinos with half spin
following a mixed statistics was not thoroughly studied earlier. For nucleons 
however such studies exist \cite{spin1}.
The recent paper on the possibility of neutrinos violating the Pauli 
Exclusion Principle [1] however has renewed interest in the 
subject \cite{Kuzmin}.
The Doglov-Smirnov work \cite{DS} points out that if 
neutrinos do obey Bose-Einstein (BE) statistics instead of Fermi-Dirac 
(FD), then 
they may form large cosmological Bose condensates and account 
for the dark 
matter. This also opens up the possibility of large lepton
asymmetry in the universe.
 As double beta decay disallows purely bosonic neutrinos, one will be
interested in mixed statistics in terms of a continuous 
``Fermi-Bose'' parameter,
$\kappa$ ($\kappa =-1$ 
is purely fermionic and $\kappa =1$ is purely bosonic) \cite{DSBBN}.
 This on the other hand has astrophysical consequences as well. For
example this will have impact on the type II supernova (SN) 
dynamics and will change the
energy spectrum of neutrinos coming out of the supernova. In this report we
shall not be concerned with the justification of the 
Dolgov-Smirnov suggestion
but concentrate on the question: Given the 
scenario of neutrinos of all three 
flavors obeying the mixed statistics, 
how would future observation of galactic 
neutrinos in large terrestrial detectors put 
limits on the mixed statistics 
parameter $\kappa$. We assume that apart from (possibly) violating
the spin-statistics theorem, the massive neutrinos do not have any 
other non-standard property.
We shall see that with detectors 
like Super-Kamiokande (SK) one indeed can test this 
hypothesis at a significant level of accuracy.

Massive stars at the end of their normal lifespan, collapse 
due to the gravitational pull
once the silicon burning in the core stops 
and the core has a mass greater
than the Chandrasekhar mass. The neutrinos that are produced by 
electron capture
on nuclei and free protons during the initial collapse phase
escape. However, as the density of the core exceeds 
densities of
$10^{11}-10^{12}$ gm/cc during collapse, neutrinos get trapped. 
At densities higher than the nuclear 
matter density a shock wave forms 
inside the core and travels outward. Whether
the shock wave can reach the edge of the 
core with enough energy to cause the
explosion with the observed energies 
is the central question of supernova 
physics today. 

At the high densities and temperatures of the SN core
during the post-bounce phase, 
neutrinos and antineutrinos of all three
flavors are produced. 
Almost $\sim 99\%$ of the 
gravitational energy produced through the huge 
contraction of the star 
($\sim$ a few
times $10^{53}$ ergs) 
is released in the form of the six species of 
neutrinos ($\nu_e$,
$\nu_{\mu}$, $\nu_{\tau}$ and their antiparticles). 
In the high density central part of the 
core, neutrinos of each flavor have
high opacity and hence cannot come out. 
But at larger radius due to a strong  
density gradient, neutrinos can diffuse out and eventually escape.
Therefore, we expect almost thermal spectra for all the neutrino 
species, with temperatures characteristic of their radius of 
last scattering in the SN, usually called the neutrino sphere radius. 
Since $\nue$ and $\anue$ have 
charged current interactions with the SN matter in addition to
neutral current interactions and since SN matter is 
neutron rich, $\nue$, $\anue$ and $\nu_x$ ('x' stands
for $\mu$, $\tau$, $\bar\mu$ and $\bar\tau$) 
decouple at different radii.
Hence the neutrino spheres of the three 
different types, $\nu_e$, 
$\bar\nu_e$ and $\nu_x$, have different  
radii and hence different equilibrium temperatures 
($T_{\nu}$), with $ T_{\nu_e} < T_{\bar\nu_e} <T_{\nu_x} (T_x)$.
For our purpose we shall take $T_\nue=3.5$ MeV, 
$T_\anue=5$ MeV and $T_{\nu_x}=8$ MeV \cite{snstandard}, used 
in most simulations and calculations as the standard values.
As there are arguments which claim that $T_\anue$ and 
$T_{\nu_x}$ should be closer \cite{snnew}, we vary these 
temperatures over a range as detailed later in this report.
These temperatures are obtained by using FD equilibrium 
energy distributions and may change somewhat had the simulations been 
carried out with BE
distributions or distributions with mixed statistics. 
However the qualitative
results of our investigation here will not change if 
they are done with equilibrium
BE temperatures for $\nu_e$, $\bar\nu_e$ and $\nu_x$. Again realistic 
simulations indicate small 
departures from equilibrium distribution of energies
of the neutrino, with the high energy tail lower or 
``pinched''. The pinching factor\footnote{The definition 
of $\eta$ is given in Eq. (\ref{Eq:eta}).} 
is expected to range between
$\eta_e \sim 0-3$, $\eta_{\bar e}\sim 0-3$ and $\eta_x \sim 0-2$ 
for the $\nue$, $\anue$ and $\nu_x$ flux respectively \cite{pinch}. 
We further assume that the total luminosity $E_B$ is equipartitioned 
among the six neutrino and antineutrino species.
We remind readers here that the 
electron type antineutrinos 
(with at most one neutrino scattering event) were
detected alongwith the SN 1987A explosion. 
But with 11 and 8 events at
Kamioka and IMB, fitting of energy spectrum and 
determination of temperature had large errors \cite{sn1987a}.

\begin{figure}
\begin{center}
\includegraphics[width=12.0cm, height=7.8cm]
{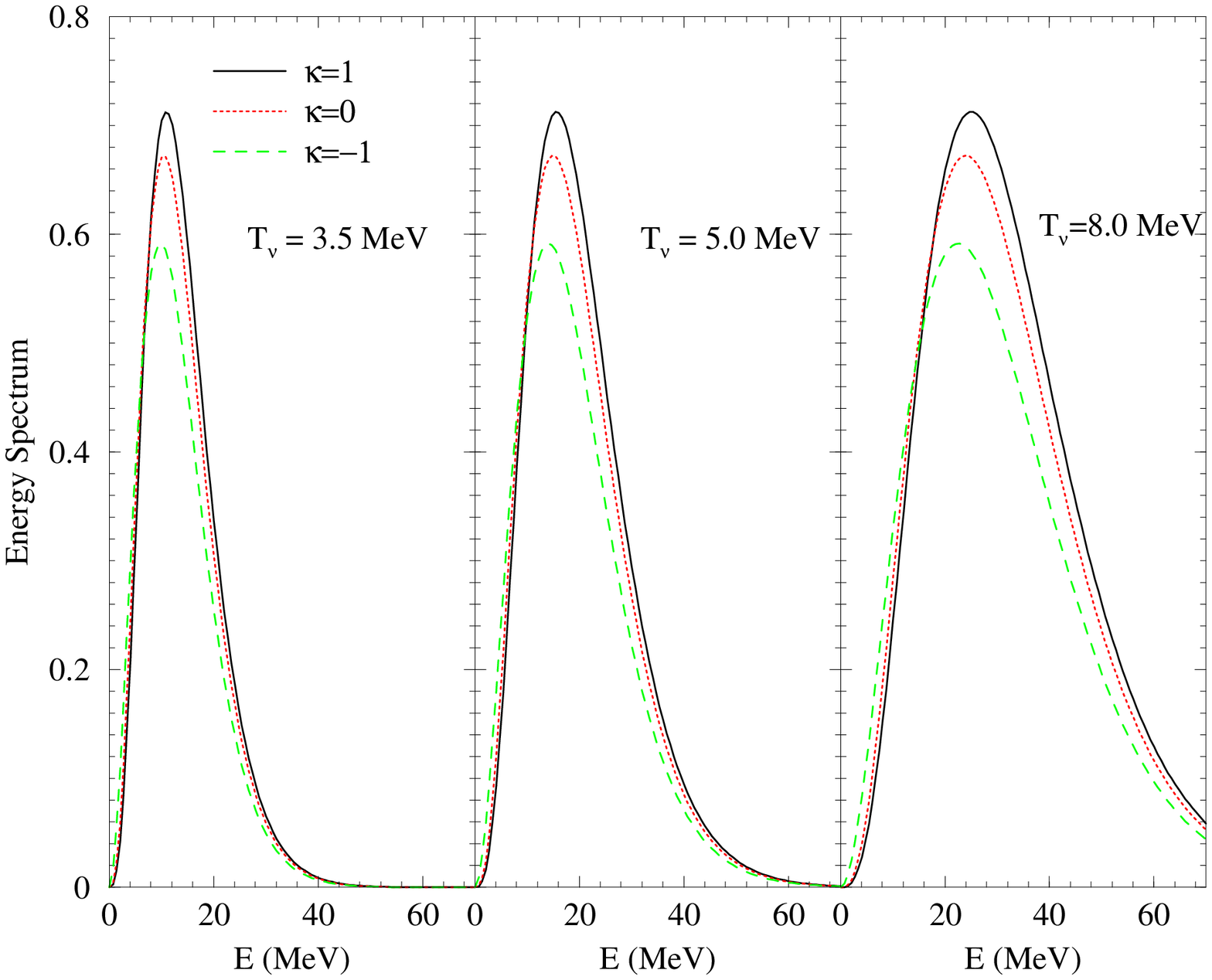}
\caption{The thermal energy spectrum for the FD, MB and BE 
distribution of SN neutrinos with 
typical temperature of 3.5 MeV (left panel), 5 MeV (middle panel)
and 8 MeV (right panel).
}
\label{fig:spectra}
\end{center}
\end{figure}

Following \cite{DS} we parametrize the
equilibrium distribution of the neutrinos with a mixed statistics
as
\be
f^{(eq)}_{\nu}  =  ({\rm exp}(E/T)+\kappa)^{-1}~,
\ee
where $E$ is the energy of the neutrinos, $T$ the temperature of the 
distribution and $\kappa$ is the Fermi-Bose parameter for the
mixed statistics, as mentioned earlier. 
Fig. \ref{fig:spectra} shows the energy spectrum of the neutrinos
corresponding to FD ($\kappa=+1$), Maxwell Boltzmann 
MB ($\kappa=0$) and BE ($\kappa=-1$) distributions, with typical
temperatures of 3.5 (left panel), 5 (middle panel) 
and 8 MeV (right panel). The area under each of the curves 
gives the average energy expected for the corresponding values of 
$T_\nu$ and $\kappa$.
It is clear from the figure that for the 
same $T_\nu$, at small values of $E$ 
the predicted spectrum for the BE distribution 
is higher than for the FD distribution. However for 
large $E$, the trend is reversed and the spectrum predicted
for FD is higher than that for the BE distribution and
has a longer high energy tail. This trend reflects the fact 
that for the same $T_\nu$, 
the predicted average energy for the neutrinos is 
larger for the FD distribution. We also stress from  
Fig. \ref{fig:spectra} the fact that the spectral shape 
for the FD and BE distributions are different. 
We want to use the signature of 
this difference in predicted average energy  
and spectral shape of the SN neutrinos
in terrestrial detectors, to 
disfavor the ``wrong'' neutrino statistical distribution 
and put limits on the Fermi-Bose parameter $\kappa$.

\begin{figure}
\begin{center}
\includegraphics[width=12.0cm, height=7.8cm]
{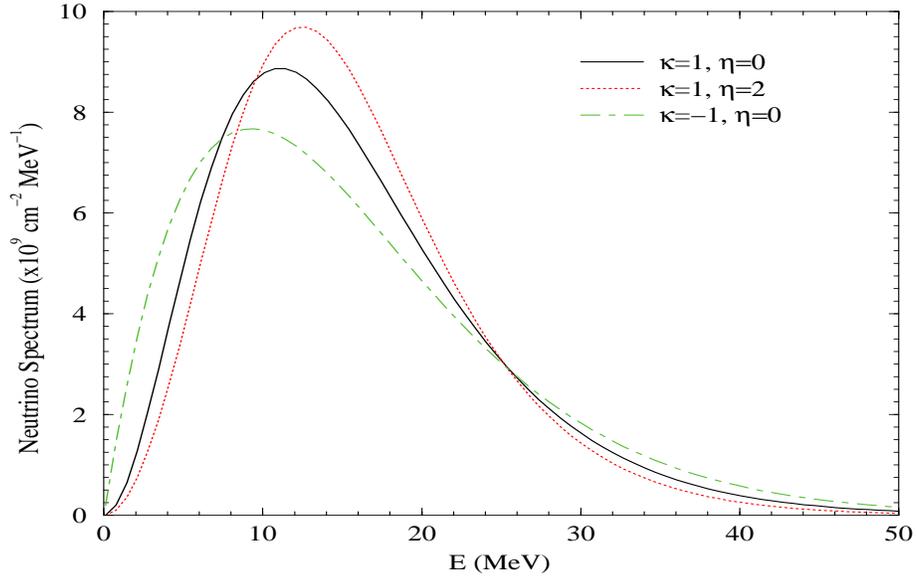}
\caption{The neutrino spectrum expected on earth from a 
galactic SN for the pure FD (black solid line), pinched FD
(red dotted line) and pure BE distribution (green dot-dashed line).
The average energy for all the three spectrum is taken as 15.75 MeV.
}
\label{fig:spec_eta}
\end{center}
\end{figure}

In Fig. \ref{fig:spec_eta} we show the effect of introducing 
the pinching in the FD distribution. We plot the 
galactic SN neutrino flux spectrum expected at earth 
given by \cite{pinch}
\be
F_\alpha^0 = \frac{L_\alpha}{4\pi D^2 T_\alpha^4 F_3(\eta_\alpha)}
\frac{E^2}{({\rm exp}(E/T_\alpha - \eta_\alpha)+\kappa)}~,
\label{Eq:eta}
\ee
where $L_\alpha$ is the SN luminosity released in $\nu_\alpha$,
$D$ is the distance of the SN (10 kpc), $\eta_\alpha$ is 
the pinching factor and 
\be
F_3(\eta_\alpha) = \int_0^\infty \frac{x^3}{{\rm exp}
(x - \eta_\alpha)+\kappa} dx~.
\ee
The black solid line shows the spectrum for a pure FD 
distribution with $\eta_\alpha=0$, the red dotted line 
corresponds to a pinched FD distribution with $\eta_\alpha=2$ while 
the green dot-dashed line is for the pure BE distribution.
All the three spectra plotted in the figure have the same 
average energy\footnote{Note that this is different from 
Fig. \ref{fig:spectra} where the temperature was kept same 
for the different distributions.}.
A comparison of the three spectra shows that the difference 
in the spectral shape between the BE and  
FD distribution increases as $\eta_\alpha$ changes from 
zero. Hence, we expect it would be easiest to confuse
between the BE and FD distributions when  $\eta_\alpha=0$ and 
it is for this case that we generally expect the minimum 
\chisq{}. Since we want to check how well one can differentiate
between the FD and BE (and mixed) distributions and since 
this seems to be most difficult for the pure FD distribution, 
we will neglect the effect of pinching in the spectrum in the 
rest of the paper for simplicity\footnote{The issue of pinching 
could get complicated in some cases,
especially for a BE spectrum with chemical potential and 
in these cases one has to be very careful. We propose to 
make a detailed study of this in a future work. However, 
we have checked that taking no pinching in the spectra yields 
the smallest \chisq{} for most realistic cases.}.

We shall be concerned with the large water Cerenkov detectors 
like SK. The main interaction channel in SK which is important
for the detection of the SN neutrinos is 
the charged current capture of $\bar\nu_e$ on free protons:
\be
\bar\nu_e + p  \rightarrow   e^+  +n 
\ee
with a threshold $\anue$ energy of 1.3 MeV. 
The number of positrons expected in SK 
from a galactic SN explosion is given by
\be
R= 
N \int dE_A \int dE_T ~R(E_A,E_T)\int dE~ 
\sigma_{\anue p} (E)~
\Big{[}F_{\bar e}(E)~ P_{\bar e\bar e}(E)
+ F_{\bar x}(E)
(1 - P_{\bar e \bar e}(E) ) \Big{]}~,
\label{anuep}
\ee
where
N is the number of target protons and
$\sigma_{\anue p}$
is the $\anue$-$p$ capture cross-sections.
$E$ is the energy of the incident $\anue$, while 
$E_A$ and $E_T$ are respectively the 
measured and true energy of the emitted
positron. The true and measured 
energy of the positron are related through 
$R(E_A,E_T)$, the Gaussian energy resolution 
function of the detector given by,
\be
R(E_A,E_T) &=&
\frac{1}{\sqrt{2\pi\sigma_0^2}}\exp\left( - \frac{(E_A - E_T)^2}
{2\sigma_0^2}   \right)~,
\ee
where we assume $\sigma_0$ to have a value similar to what is 
used in the analysis of the SK solar neutrino data. 
$F_{\bar e}$ and $F_{\bar x}$ are the flux of $\anue$ and $\bar\nu_x$ 
produced in the SN core 
and depend on the spin statistics of the neutrinos
and $P_{\bar e\bar e}$ is the survival probability of the SN $\anue$.

Before reaching the detector, the neutrinos 
travel through the SN matter, 
propagate through vacuum and finally may 
even travel through earth matter depending 
on the time of the SN burst. For
the range of neutrino oscillation parameters consistent with 
the solar and KamLAND reactor data 
($\Delta m^2_{\odot} \equiv \Delta m^2_{21}=8.0 \times 
10^{-5}$ eV$^2$ and 
$ \sin^2 \theta_{\odot} \equiv \sin^2 \theta_{12}=0.31$ 
\cite{solar,solarus}) 
and SK atmospheric data ($\Delta m^2_{\rm atm} 
\equiv \Delta m^2_{31}=2.1 \times 10^{-3}$ eV$^2$ and
$ \sin^2 2\theta_{\rm atm}  \equiv \sin^22\theta_{23} =1.0$ \cite{skatm}), 
one expects large matter enhanced oscillation inside the 
supernova and 
possible regeneration effects inside the earth 
\cite{pinch,smirnov}. 
As a result,
the more energetic $\bar\nu_{\mu}$ and $\bar\nu_{\tau}$ get converted to 
$\bar\nu_e$ and make the spectra harder. 
This results in enhancing the number of expected charged current 
events in the detector and is incorporated in our numerical 
analysis through $P_{\bar e \bar e}$ in Eq. (\ref{anuep}).
We assume that the 
supernova explodes at a distance of 10 kpc from us 
and that the time of the event is such that the 
neutrinos do not cross the earth matter. The 
survival probability $P_{\bar e \bar e}$ depends critically 
on the value of $\theta_{13}$, which at the moment has a very 
weak bound of $\sch < 0.04$ at \sig{} \cite{solar,chooz}.
In what follows, we will assume that the true values of the 
oscillation parameters $\ms$, $\sss$ and $\ma$ have been determined 
with sufficient accuracy to the values mentioned above 
(see \cite{futuresol} and 
\cite{futureatm}
for recent detailed discussions). 
For the major part of the paper, we will also assume that the 
true value of mixing angle $\theta_{13} \sim 0$. We will discuss 
the impact of the uncertainty on this oscillation parameter
towards the end.

\begin{figure}[b]
\begin{center}
\includegraphics[width=12.0cm, height=7.8cm]
{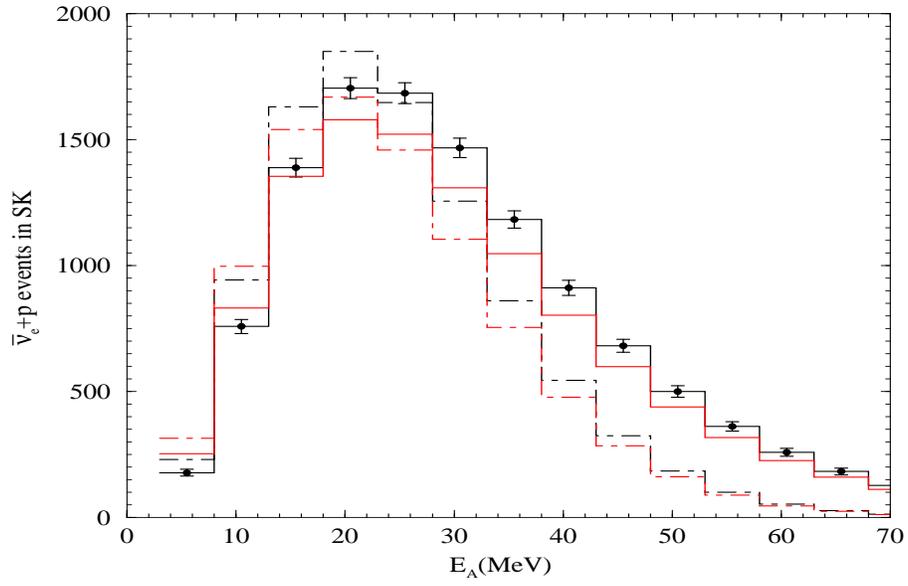}
\caption{The $(\anue+p)$ event spectrum 
expected in SK from a galactic SN. The solid 
black and red lines 
show the events spectrum for FD and BE distribution 
respectively where neutrino oscillations have been taken into 
account. The dashed black and red lines 
show the corresponding spectra without oscillations.
}
\label{fig:events}
\end{center}
\end{figure}

Fig. \ref{fig:events} displays the distribution of 
the number of $ (\bar\nu_e + p)$ events in SK as a
function of the measured energy $E_A$ of the positron.
The black solid (dashed) line gives the event spectrum  
for a FD distribution with (without) neutrinos oscillations,
while the red solid (dashed) line gives the expected event 
spectrum for a BE distribution. We assume 
$T_\anue=5$ MeV, $T_x=8$ MeV and $E_B=3\times 10^{53}$ ergs.
From the figure, BE distribution 
can be seen to underpredict the number of events 
over a large range of energy from about 12-70 MeV, 
compared to FD distribution.
We also give the $\pm 1\sigma$ error bars corresponding to the 
number of events SK would observe in each bin for the FD distribution. 
This clearly shows that a large detector like 
SK should be able to distinguish between the different 
statistics at a good confidence level.

In order to quantify our statement, 
we perform a statistical analysis of the neutrino induced 
positron spectrum that SK would observe in the event of 
a galactic SN explosion. For the errors we take a Gaussian
distribution and define our \chisq{} as
\be
\chi^2 = \sum_{i,j}(N_i^{theory}-N_i^{obsvd})(\sigma_{ij}^2)^{-1}
(N_j^{theory}-N_j^{obsvd})~,
\label{eq:chi}
\ee
where $N_i^{theory}$ ($N_i^{obsvd}$) are the number of predicted
(observed) events in the $i^{th}$ energy bin and the sum is over 
all bins. The error matrix
$\sigma_{ij}^2$ contains the statistical and systematic errors. 
The systematic uncertainty will be eventually determined by the 
experimental collaboration after a real SN event has been observed 
in SK. However, for the moment we have simply assumed a 5\%\footnote
{Note that this is a very conservative estimate. The total 
systematic error in the SK solar neutrino data sample is only
3.5\% \cite{sksolarfinal}.} 
systematic uncertainty in the simulated ``data'',
fully correlated in all the bins. The number of energy bins 
used in our analysis is shown in Fig. \ref{fig:events}.
The \chisq{} function 
is minimized with respect to the parameters in 
the theory to give us a handle on the C.L. with which 
observation of SN neutrinos in SK could lead to the determination 
of their correct distribution function. In what follows, 
there are two kinds of statistical 
tests that we shall perform. In one, we shall compare the data set 
obtained for (say) the true FD distribution with the predicted 
event spectrum for the false BE distribution. This will give us 
the measure of the C.L. with which the false BE distribution  
could be ruled out. In another, we will compare the data set
generated for (say) the true FD distribution ($\kappa=1$)
and compare 
it with a distribution for all possible values of $\kappa$ 
in the range [$-1,1$], with $\kappa=-1$ corresponding to a pure
BE distribution.

\begin{figure}
\begin{center}
\includegraphics[width=12.0cm, height=7.8cm]
{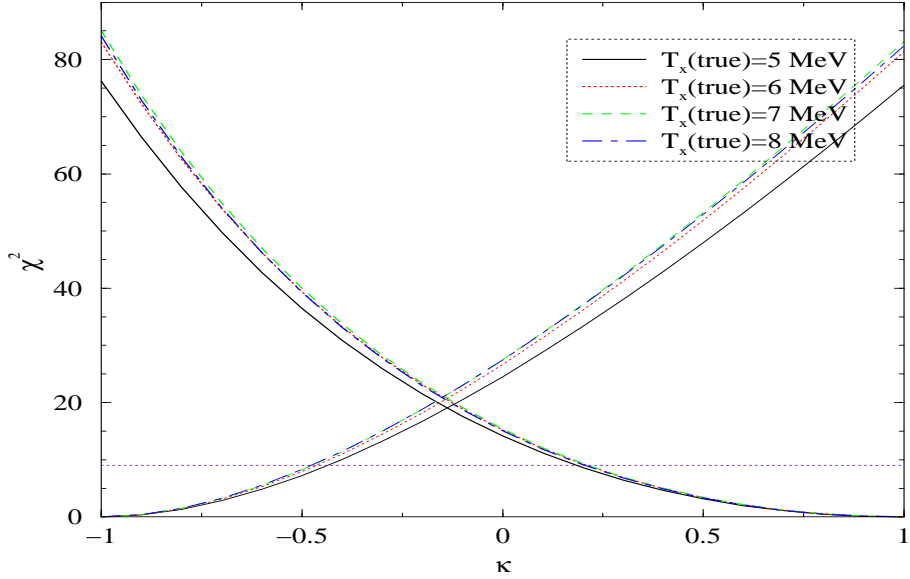}
\caption{The \chisq{} as a function of $\kappa$.
The thick (thin) lines show the \chisq{} obtained 
between a data set for a true 
FD (BE) spectrum and a theoretical spectrum with mixed statistics.
The four lines correspond to 
four different values of $\txt$. For all cases we have taken a 
fixed value of $\taet=5$ MeV. The oscillation parameters and 
$E_B$ are also kept fixed.
}
\label{fig:chikappa}
\vskip -0.5cm
\end{center}
\end{figure}

In Fig. \ref{fig:chikappa} we show the $\chi^2$ as a function of the 
parameter $\kappa$. We assume that the only errors involved are
experimental and keep all the theoretical inputs fixed. 
The thick lines show the \chisq{} for the case
where the $N_i^{obsvd}$ correspond to $\kappa{\rm (true)}=1$ 
and hence to a 
``true'' FD spectrum. The thin lines give the corresponding $\chi^2$ 
for the case when $\kappa{\rm (true)}=-1$ and the BE spectrum is true.
For all the cases we generated the ``data'' for $\taet=5$ MeV, while for 
$\txt$ we assume four different values displayed in the figure. 
If we define\footnote{Throughout this paper we will define 
our $n\sigma$ limit as given by $\Delta \chi^2=n^2$. Note that 
this is just a definition we {\it assume} 
here for the sake of comparison of the C.L. we obtain
for the different types of fits. The C.L. in general should depend
on the number of constrained degrees of freedom in the theory.}
our \sig{} limit as given by $\Delta \chi^2=9$, then 
the figure shows that for a true FD spectrum, 
SK could in principle restrict $\kappa > 0.2$ at \sig, for
$\taet=5$ MeV and $\txt=8$ MeV\footnote{There is very little 
dependence on the value of $T_x$.}. 
In particular, the false BE 
distribution ($\kappa=-1$)
can be ruled out comprehensively at about $9\sigma$ level. 
If on the other hand BE distribution was true,
then all $\kappa > -0.5$ could be ruled out at \sig{} and the 
false FD distribution ($\kappa=1$) could be disfavored at 
about $9\sigma$ level.

In the discussion above we had assumed that all theoretical 
parameters involving the prediction of the SN neutrino fluxes were 
absolutely known. However, this is far from true. In fact, 
the error coming from our lack of 
correct modeling for the supernova neutrino energies and luminosities
is much larger than the experimental errors. 
In the last few years it has been realized \cite{snnew} that the nucleon
bremsstrahlung $NN\rightarrow NN\nu_x\bar\nu_x$ as well as 
$\nue\anue \rightarrow \nu_x\bar\nu_x$  along with its inverse 
reaction are important and were not included in the 
earlier simulations of the neutrino transport inside the supernova.
While the first process 
reduces the difference between $T_x$ and $T_\anue$,
the second ones increase the $\nu_x$ luminosity. 
We note that these uncertainties could also be correlated.
In what follows, we 
take into account the uncertainties on the 
predicted neutrino fluxes by keeping
both the total luminosity $E_B$ and 
the temperatures $T_\anue$ and $T_x$ completely unconstrained,
thus allowing for all possible values for these parameters.
This method therefore can be used to gauge the potential of 
SK to determine the true energy distribution of the SN neutrinos,
irrespective of our (lack of) knowledge of the initial 
SN fluxes. Henceforth, we will distinguish between the ``true'' 
values of the parameters (denoted as $\taet$ etc.)
at which we generate our projected SN 
neutrino data set and the fitted values of those parameters
(denoted as $T_\anue$ etc.), obtained 
through the minimization of the \chisq{} function defined in 
Eq. (\ref{eq:chi}).

\begin{figure}
\begin{center}
\vskip -0.5cm
\includegraphics[width=12.0cm, height=7.8cm]
{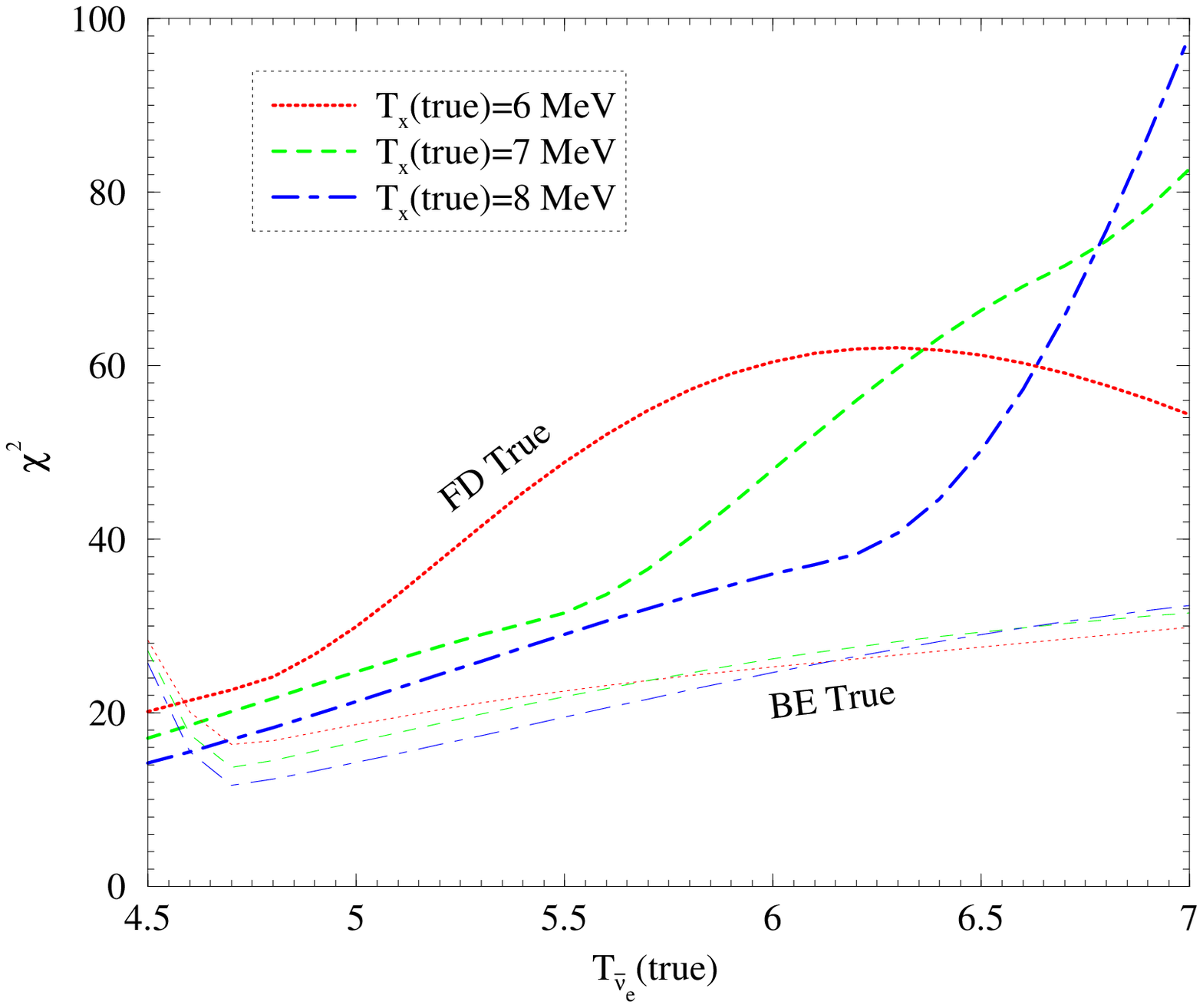}
\caption{
The \chisq{} as a function of $\taet$ 
between a data set obtained for a true 
FD/BE spectrum and a theoretical spectrum with BE/FD statistics.
The thick lines are for data set with true FD statistics and 
the thin lines for data set with true BE statistics.
We take three different values for $\txt$ and let all SN parameters 
to vary freely in the fit. 
}
\label{fig:chitanueallfree}
\end{center}
\end{figure}

Fig. \ref{fig:chitanueallfree} shows the \chisq{} as a 
function of $\taet$ for three assumed values of $\txt$.
The thick lines in the figure show the case where the ``data''
$N_i^{obsvd}$ is generated assuming a true FD distribution
for $E_B{\rm (true)}=3\times 10^{53}$ ergs and
values of $\taet$ between 4.5-7 MeV 
and for three different values of 
$\txt$ = 6, 7 and 8 MeV shown by red dotted and green 
dashed and blue dot-dashed lines
respectively. This data set is then fitted with 
$N_i^{theory}$ calculated using 
the wrong BE distribution and with 
$T_\anue$, $T_x$ and $E_B$ allowed to take any possible 
value which gives the minimum \chisq{} 
in the fit. The thin lines in the 
figure show the corresponding minimum \chisq{} when the true 
distribution appearing in $N_i^{obsvd}$
is BE and the wrong distribution chosen in $N_i^{theory}$ is 
FD. We note from the figure that if FD 
is the true distribution, 
then the SN data set in SK can be used to rule out the possibility 
of neutrinos obeying the BE 
statistics at the $5\sigma$ level 
for $\taet > 4.8$, 5.0 and 5.2 MeV for $\txt=6$, 7 and 8 MeV
respectively.
If the BE 
distribution for the neutrinos was true, then 
the wrong FD 
statistics could be ruled out at $5\sigma$ for 
$\taet > 6.0$, 5.9 and 6.1 MeV, when $\txt=6$, 7 and 8 MeV
respectively. The wrong distribution can be ruled out at 
more than \sig{}
for both cases for all plausible values of $T_\anue$.

\begin{figure}
\begin{center}
\includegraphics[width=12.0cm, height=8cm]
{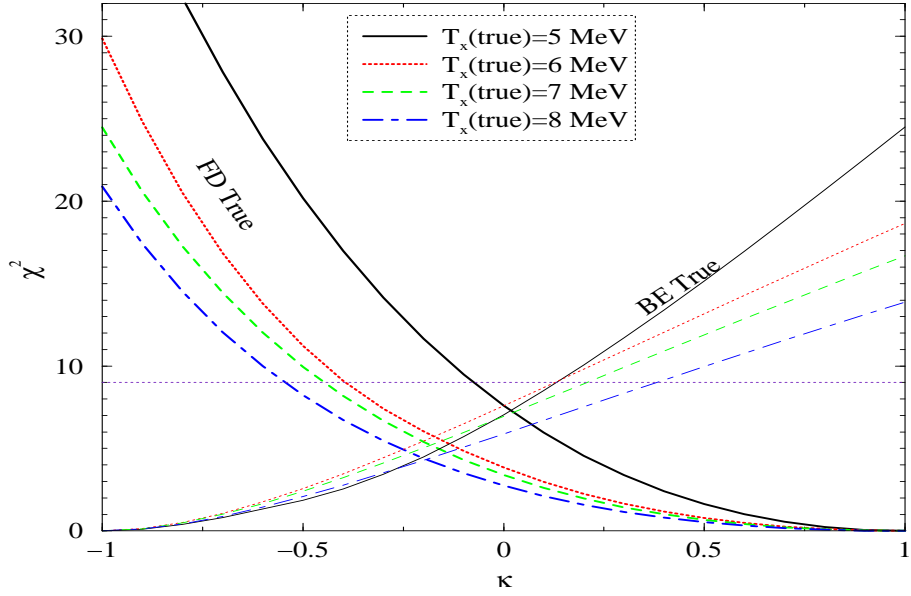}
\caption{
The \chisq{} as a function of $\kappa$
between a data set obtained for a true 
FD/BE spectrum and a theoretical spectrum with mixed statistics.
The thick lines are for data set with true FD statistics and 
the thin lines for data set with true BE statistics.
We take four different values for $\txt$ and let all SN parameters 
to vary freely in the fit. 
}
\label{fig:chikappaallfree}
\end{center}
\end{figure}

The C.L. with which any given value of $\kappa$ can be 
ruled out for either the case where FD (thick lines)
or BE (thin lines) is the true distribution, 
can be seen from Fig. \ref{fig:chikappaallfree}, where we have 
plotted the \chisq{} as a function of $\kappa$. For all the curves
we take $\taet=5$ MeV, while for $\txt$ we have assumed four 
values: 5 (black solid lines), 6 (red dotted lines), 
7 (green dashed lines) and 8 (blue dot-dashed lines)
MeV. The method of $\chi^2$ analysis is 
similar to that used for Fig. \ref{fig:chitanueallfree}. 
In particular, for the case of true FD (BE) 
distribution shown by the 
thick (thin) lines, the data set $N_i^{obsvd}$ is generated for 
$\kappa{\rm (true)}=1$ ($\kappa{\rm (true)}=-1$), 
$E_B{\rm (true)}=3\times 10^{53}$ 
and the fixed values of $\taet$ and 
$\txt$ mentioned above. This data set 
is then fitted with $N_i^{theory}$
calculated for each $\kappa$ in the range [$-1,1$], with
the SN parameters allowed to vary freely. 
For the true FD distribution, we can limit allowed
$\kappa > -0.39 $, $-0.45$ and $-0.55$ at \sig, for 
$\txt=6$, 7 and 8 MeV respectively. For BE as the true 
distribution, $\kappa < 0.13 $, 0.21 and 0.38 is allowed at \sig, for 
$\txt=6$, 7 and 8 MeV respectively.

From the Fig. \ref{fig:chikappaallfree} we note that when a 
theoretical distribution with a wrong $\kappa$
is used to fit the data, the $\chi^2$ obtained for
the true FD distribution is in general larger than 
that for the true BE distribution. This happens because 
by changing the (anti)neutrino temperatures and luminosity, 
a spectrum with mixed statistics can reproduce 
the data set more easily when the latter corresponds to a 
BE distribution. We have checked that the spectral shape 
is the most important factor in this case. 
Indeed if we analyzed the data set on total $(\anue+p)$
observed rate in SK rather 
than the positron energy spectrum, the sensitivity to $\kappa$
would get completely lost since the total rate can always be 
reproduced by a theory with any $\kappa$ 
when $E_B$, $T_\anue$ and $T_x$ are allowed to vary freely.

\begin{figure}
\begin{center}
\includegraphics[width=12.0cm, height=8cm]
{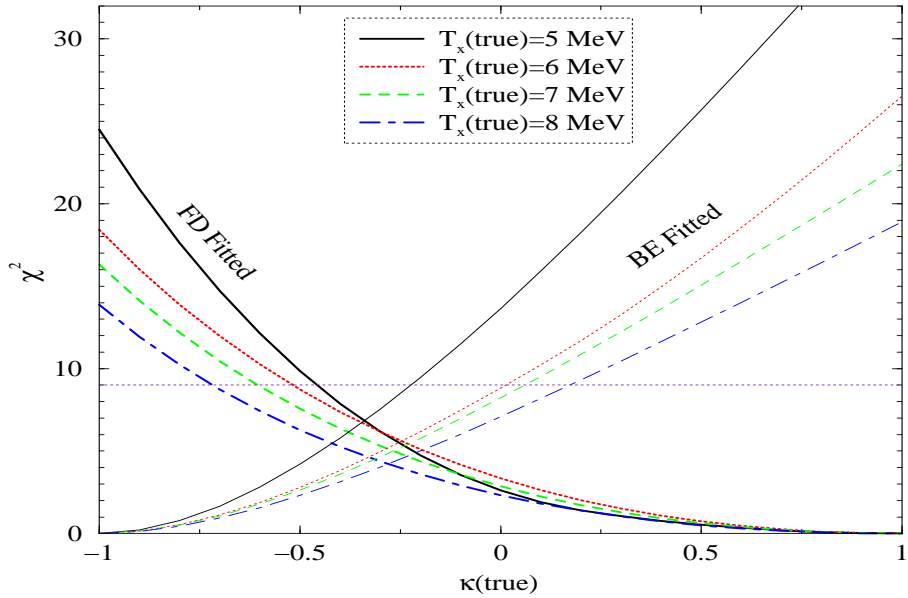}
\caption{The \chisq{} as a function of $\kappa{\rm (true)}$.
Everything else is same as in Fig. \ref{fig:chikappaallfree}.
}
\label{fig:chikappatrueallfree}
\end{center}
\end{figure}

Fig. \ref{fig:chikappatrueallfree} shows the $\chi^2$ 
we can expect when a data set corresponding to a particular case of 
mixed neutrino spin-statistics is fitted with either a pure FD
or a pure BE distribution.
We generate the data set corresponding to each of 
values of $-1 < \kappa({\rm true}) <1$ and fit it 
with either a theory with FD distribution (thick lines) or 
a theory with BE distribution (thin lines). We can draw 
similar inferences from this plot as obtained from 
Fig. \ref{fig:chikappaallfree}.

Throughout our discussion so far we have played down the impact 
of the oscillation parameters. We generated our data set at the 
benchmark values of 
$\ms=8\times 10^{-5}$ eV$^2$, $\sss=0.3$, 
$\ma=2.1\times 10^{-3}$ eV$^2$ and $\sch=0$ and kept
them fixed in the fit.  
For the current allowed range of neutrino oscillation 
parameters, 
the survival probability for the SN (anti)neutrinos 
depend on $\ms$ and $\sss$ only for (anti)neutrinos crossing the earth.
Since we assume that the position and time 
of the SN in the galactic center
with respect to our detector SK will be 
such that neutrinos do not cross the earth matter, we are justified 
in keeping $\ms$ and $\sss$ fixed at any value of their current 
allowed range. However, the oscillations inside the SN matter depend 
crucially on the values of $\ma$ and $\theta_{13}$ \cite{smirnov}. 
Most importantly, for a given value of $\theta_{13}$ ($>10^{-6}$), 
whether large flavor oscillations appear in the 
neutrino or the antineutrino channel is determined by the 
sign of $\ma$, {\it i.e.} the neutrino mass hierarchy. 
The sign of $\ma$ is typically expected to be determined using 
matter effects in the 1-3 channel. 
For very large values of $\theta_{13}$, synergies between the 
T2K and NO$\nu$A experiments could somewhat indicate the 
$sgn(\ma)$ \cite{synergy}. 
However, an unambiguous measurement would require 
either a beta beam 
facility or a neutrino factory \cite{nufac}. 
Resonant matter effects in the 1-3 channel 
encountered by atmospheric neutrinos can be exploited to 
probe the neutrino hierarchy both in water Cerenkov and large 
magnetized iron calorimeter detectors \cite{hie_atm}. 
Recently some 
novel ways of probing the mass hierarchy requiring very precise 
measurements and using the ``interference
terms'' between the different oscillation frequencies have been proposed
\cite{way_out_hie}. 
Neutrinoless double beta decay experiments have the potential to 
provide us with the neutrino mass hierarchy even for vanishing 
$\theta_{13}$ \cite{cr0vbb,0vbbothers}.
The uncertainty on the value/limit on $\theta_{13}$ is also 
expected to reduce in the future. 
The upper limit could be improved further to 
$\sch \ltap 0.0025$ \cite{huber10} by 
the combination of the next generation 
beam experiments T2K and NO$\nu$A,
as well as by the second generation 
reactor experiments \cite{react13}. 
It should be 
borne in mind that any one of these experiments could 
even measure a nonzero $\theta_{13}$, if the true value 
of $\theta_{13}$ happens to fall within their range of sensitivity.

\begin{figure}
\begin{center}
\includegraphics[width=12.0cm, height=8.5cm]
{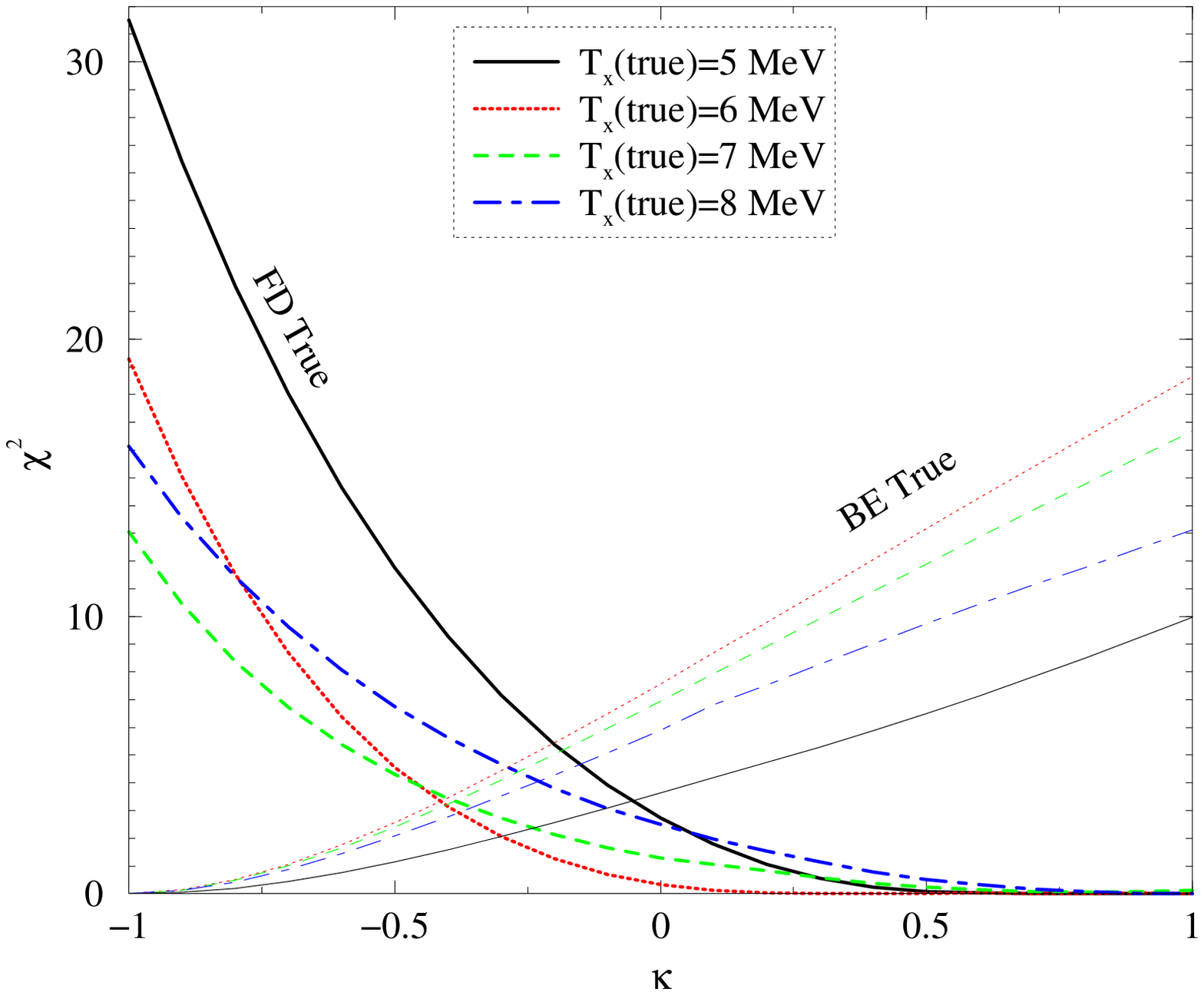}
\caption{Same as Fig. \ref{fig:chikappaallfree} but with even 
$\theta_{13}$ allowed to vary freely in the fit.
}
\label{fig:chikappat13free}
\end{center}
\end{figure}

In our analyzes so far, we had kept $\theta_{13}$ 
fixed at zero. For this case the oscillation probability is
independent of the neutrino mass hierarchy 
for neutrinos not crossing the earth. Hence, our 
study so far is valid irrespective of the neutrino mass hierarchy.
However, there is no reason to assume that the mixing angle 
$\theta_{13}=0$. In fact, large number of theoretically attractive 
models predict that $\theta_{13}$ is large and possibly close to its 
current upper limit.
If the true value of $\theta_{13}$ is indeed large then we expect 
big matter induced flavor oscillations in the SN for the 
antineutrino (neutrino) channel, when the neutrino mass hierarchy
is inverted (normal). For $10^{-6} \ltap \sch \ltap 10^{-4}$ these 
matter induced conversions in the antineutrino (neutrino) channel
for inverted (normal) hierarchy are also energy dependent,
causing larger suppression for antineutrinos (neutrinos) 
with smaller energies \cite{smirnov}. 
To take into account the impact of 
our lack of knowledge of the true value of $\theta_{13}$ we present 
the \chisq{} as a function of $\kappa$ in 
Fig. \ref{fig:chikappat13free}, which is similar to 
Fig. \ref{fig:chikappaallfree} but with $\theta_{13}$ 
allowed to vary freely as well.
The survival probability for $\anue$ 
is same for both normal and inverted hierarchy 
for $\sch \ltap 10^{-6}$.
For the normal hierarchy
the survival probability for $\anue$ remains the same for all
$\theta_{13}$, while for the inverted hierarchy the 
probability reduces as $\sch$ increases. Therefore, 
by taking the inverted hierarchy in the fit and 
allowing the value of $\sch$ to vary in the full range [0-0.04],
we take into account both the uncertainty due to $\theta_{13}$ as 
well as the hierarchy \cite{lunardini05}.
A comparison of Fig. \ref{fig:chikappat13free} with 
Fig. \ref{fig:chikappaallfree} brings out the fact that 
the $\chi^2$ for the case when the data set corresponds to FD spectrum 
reduces substantially when $\theta_{13}$ is allowed to vary in the fit.
As discussed in detail before, the SN neutrino spectrum depends on 
whether the neutrinos follow the FD, BE or mixed statistics. 
In particular, we saw in Figs. \ref{fig:spectra} and 
\ref{fig:events} that the FD 
distribution allows for a spectrum enriched in higher energy 
and depleted in lower energy (anti)neutrinos, 
compared to a spectrum with BE or mixed statistics.
Since as noted above, for $10^{-6} < \sch < 10^{-4}$ there is 
larger suppression in the lower energy end of the event spectrum
due to matter induced oscillations, a theoretical BE (or mixed
statistics) spectrum with 
$\sch$ in the range [$10^{-6}$-$10^{-4}$] and with inverted hierarchy
can be reconciled better with a 
data set generated for the FD distribution with $\theta_{13}=0$.
This helps in reducing the \chisq{} when the FD distribution is true.
However, we can see from the Fig. \ref{fig:chikappat13free}
that the \chisq{} corresponding to the true BE distribution 
does not change much due to the uncertainty in $\theta_{13}$ and 
hierarchy. This is because oscillations further 
accentuate the problem if a true BE distribution is to be 
fitted with a $\kappa > -1$, except when $\taet \approx \txt$.

To summarize, we use the number of positron events induced in SK 
through the $\anue$ capture on protons for a future 
type-II galactic supernova,
to probe whether the spin-statistics is violated for 
neutrinos. We assumed that the neutrinos 
behave as prescribed by the standard model (including massive 
neutrinos) in all respects, 
except for the fact that their thermal distribution might 
obey a mixed statistical distribution rather than FD.
We performed a detailed $\chi^2$ analysis of the ``observed'' 
event spectrum in SK to find the allowed range of the Fermi-Bose 
parameter $\kappa$. We have presented results incorporating 
the uncertainties stemming from 
the astrophysical parameters for the supernova as well as 
those coming from neutrino oscillation parameters.

\vskip 1cm
\noindent{\bf Acknowledgement:} The authors wish to thank 
D. Bandyopadhyay for discussions.



\newpage


\begin{thebibliography}{}

\bibitem{DS}
  A.~D.~Dolgov and A.~Y.~Smirnov,
  Phys.\ Lett.\ B {\bf 621}, 1 (2005).

\bibitem{spin1}
  A.~Y.~Ignatiev and V.~A.~Kuzmin,
Yad. Fiz. {\bf 46}, 786 (1987) [Sov. J. Nucl. Phys. {\bf 46}, 786 (1987)];
  JETP Lett.\  {\bf 47}, 4 (1988).

\bibitem{spin2}
  L.~B.~Okun,
  JETP Lett.\  {\bf 46}, 529 (1987).

\bibitem{spin3}
  O.~W.~Greenberg and R.~N.~Mohapatra,
  Phys.\ Rev.\ Lett.\  {\bf 59}, 2507 (1987)
  [Erratum-ibid.\  {\bf 61}, 1432 (1988)];
%
  O.~W.~Greenberg and R.~N.~Mohapatra,
  Phys.\ Rev.\ Lett.\  {\bf 62}, 712 (1989)
  [Erratum-ibid.\  {\bf 62}, 1927 (1989)];
%
  O.~W.~Greenberg and R.~N.~Mohapatra,
  Phys.\ Rev.\ D {\bf 39}, 2032 (1989).

\bibitem{spin4}
  A.~B.~Govorkov,
  Phys.\ Lett.\ A {\bf 137}, 7 (1989).

\bibitem{Kuzmin} 
  A.~Y.~Ignatiev and V.~A.~Kuzmin,
  arXiv:hep-ph/0510209.


\bibitem{DSBBN}
  A.~D.~Dolgov, S.~H.~Hansen and A.~Y.~Smirnov,
  JCAP {\bf 0506}, 004 (2005).

\bibitem{snstandard}
  Y.~Z.~Qian {\it et al.},
  Phys.\ Rev.\ Lett.\  {\bf 71}, 1965 (1993).

\bibitem{snnew}
  S.~Hannestad and G.~Raffelt,
  Astrophys.\ J.\  {\bf 507}, 339 (1998);
%
  R.~Buras {\it et al.},
  Astrophys.\ J.\  {\bf 587}, 320 (2003);
%
  M.~T.~Keil, G.~G.~Raffelt and H.~T.~Janka,
  Astrophys.\ J.\  {\bf 590}, 971 (2003);
%
  M.~Liebendoerfer {\it et al.},
  Astrophys.\ J.\  {\bf 620}, 840 (2005).
%

\bibitem{pinch}
  C.~Lunardini and A.~Y.~Smirnov,
  JCAP {\bf 0306}, 009 (2003);

\bibitem{sn1987a}
A. Burrows and J.M. Lattimer, Ap. J. {\bf 318}, L63 (1987);
K. Kar in Supernova and Stellar Evolution (ed. A. Ray and T.
Velusamy; World Scientific; 1991) p 222.


\bibitem{solar}
  T.~Araki {\it et al.},  
[KamLAND Collaboration],
  Phys.\ Rev.\ Lett.\  {\bf 94}, 081801 (2005);
%
  B.~Aharmim {\it et al.},  
[SNO Collaboration],
  arXiv:nucl-ex/0502021;
%
S.~Fukuda {\it et al.},  
[Super-Kamiokande Collaboration],
Phys.\ Lett.\ B {\bf 539}, 179 (2002);
%
%
%
B.~T.~Cleveland {\it et al.},
Astrophys.\ J.\  {\bf 496}, 505 (1998);
%
J.~N.~Abdurashitov {\it et al.},  
[SAGE Collaboration],
J.\ Exp.\ Theor.\ Phys.\  {\bf 95}, 181 (2002);
[Zh.\ Eksp.\ Teor.\ Fiz.\  {\bf 122}, 211 (2002)]
[arXiv:astro-ph/0204245];
%
W.~Hampel {\it et al.},  
[GALLEX Collaboration],
Phys.\ Lett.\ B {\bf 447}, 127 (1999).

\bibitem{solarus}
  A.~Bandyopadhyay {\it et al.},
  Phys.\ Lett.\ B {\bf 608}, 115 (2005);
%
  S.~Goswami, A.~Bandyopadhyay and S.~Choubey,
  Nucl.\ Phys.\ Proc.\ Suppl.\  {\bf 143}, 121 (2005).

\bibitem{skatm}
  Y.~Ashie {\it et al.},  
  Phys.\ Rev.\ D {\bf 71}, 112005 (2005).


\bibitem{smirnov}
  A.~S.~Dighe and A.~Y.~Smirnov,
  Phys.\ Rev.\ D {\bf 62}, 033007 (2000);
%
  C.~Lunardini and A.~Y.~Smirnov,
  Nucl.\ Phys.\ B {\bf 616}, 307 (2001);
%
%
 A.~S.~Dighe, M.~T.~Keil and G.~G.~Raffelt,
  JCAP {\bf 0306}, 005 (2003);
%
  A.~S.~Dighe {\it et al.},  
  JCAP {\bf 0401}, 004 (2004);
%
  H.~Minakata {\it et al.},  
  Phys.\ Lett.\ B {\bf 542}, 239 (2002);
%
  A.~Bandyopadhyay {\it et al.},  
  arXiv:hep-ph/0312315;
%
  S.~Choubey and K.~Kar,
  arXiv:hep-ph/0212326.


\bibitem{chooz}
M.~Apollonio {\it et al.},
Eur.\ Phys.\ J.\ C {\bf 27}, 331 (2003).

\bibitem{futuresol}
  A.~Bandyopadhyay {\it et al.},
  Phys.\ Rev.\ D {\bf 72}, 033013 (2005)
and references therein.

\bibitem{futureatm}
  P.~Huber {\it et al.},
  Phys.\ Rev.\ D {\bf 70}, 073014 (2004).


\bibitem{sksolarfinal}
  J.~Hosaka {\it et al.}  [Super-Kamkiokande Collaboration],
  arXiv:hep-ex/0508053.

\bibitem{synergy}
  P.~Huber, M.~Lindner and W.~Winter,
  Nucl.\ Phys.\ B {\bf 654}, 3 (2003).

\bibitem{nufac}C.~Albright {\it et al.}  
[Neutrino Factory/Muon Collider Collaboration],
physics/0411123.

\bibitem{hie_atm} 
J.~Bernabeu {\it et al.},  
  Nucl.\ Phys.\ B {\bf 669}, 255 (2003); 
S.~Palomares-Ruiz and S.~T.~Petcov,
  Nucl.\ Phys.\ B {\bf 712}, 392 (2005);
%
  P.~Huber, M.~Maltoni and T.~Schwetz,
  Phys.\ Rev.\ D {\bf 71}, 053006 (2005); 
%
R.~Gandhi {\it et al},
hep-ph/0411252;
%
  D.~Indumathi and M.~V.~N.~Murthy,
  Phys.\ Rev.\ D {\bf 71}, 013001 (2005);
%
  R.~Gandhi {\it et al},
  arXiv:hep-ph/0506145.

\bibitem{way_out_hie}
A.~de Gouvea, J.~Jenkins and B.~Kayser,
  hep-ph/0503079;
%
H.~Nunokawa, S.~Parke and R.~Z.~Funchal,
  hep-ph/0503283;
%
S.~Choubey, S.~T.~Petcov and M.~Piai,
  Phys.\ Rev.\ D {\bf 68}, 113006 (2003); 
%
  A.~de Gouvea and W.~Winter,
  arXiv:hep-ph/0509359.

\bibitem{cr0vbb}
  S.~Choubey and W.~Rodejohann,
  Phys.\ Rev.\ D {\bf 72}, 033016 (2005)
and references therein.

\bibitem{0vbbothers}
  S.~Pascoli, S.~T.~Petcov and T.~Schwetz,
  Nucl.\ Phys.\ B {\bf 734}, 24 (2006);
%
  A.~de Gouvea and J.~Jenkins,
  arXiv:hep-ph/0507021.


\bibitem{huber10}
  P.~Huber {\it et al.}, 
  Phys.\ Rev.\ D {\bf 70}, 073014 (2004).

\bibitem{react13}
  K.~Anderson {\it et al.},
  hep-ex/0402041.

\bibitem{lunardini05}
  C.~Lunardini,
  arXiv:astro-ph/0509233.


\end{thebibliography}
\end{document}